\documentclass[letterpaper,english,reprint, aps]{revtex4-1}
\usepackage[T1]{fontenc}
\usepackage[latin9]{inputenc}
\usepackage{mathrsfs}
\usepackage{bm}
\usepackage{amsmath}
\usepackage{amssymb}
\usepackage{graphicx}

\makeatletter

\pdfpageheight\paperheight
\pdfpagewidth\paperwidth

\usepackage{cancel}

\makeatother

\usepackage{babel}
\begin{document}
\preprint{APS/123-QED}
\title{Fully kinetic simulations of strong steady-state collisional planar
plasma shocks}
\author{S. E. Anderson}
\email{andeste@lanl.gov}

\author{L. Chac\'{o}n}
\author{W. T. Taitano}
\author{A. N. Simakov}
\author{B. D. Keenan}
\affiliation{Los Alamos National Laboratory, Los Alamos, NM 87545, USA}
\date{\today}
\begin{abstract}
We report on the first steady-state simulations of strong plasma shocks
with fully kinetic ions \emph{and }electrons, independently confirmed
by two fully kinetic codes (an Eulerian continuum and a Lagrangian
particle-in-cell). While kinetic electrons do not fundamentally change
the shock structure as compared with fluid electrons, we find an appreciable
rearrangement of the preheat layer, associated with nonlocal electron
heat transport effects. The electron heat flux profile qualitatively
agrees between kinetic and fluid electron models, suggesting a certain
level of \textquotedblleft stiffness\textquotedblright , though substantial
nonlocality is observed in the kinetic heat flux. We also find good
agreement with nonlocal electron heat-flux closures proposed in the
literature. Finally, in contrast to the classical hydrodynamic picture,
we find a significant collapse in the `precursor' electric-field shock
at the preheat layer edge, which correlates with the electron-temperature
gradient relaxation.

\begin{description}
\item [{PACS~numbers}] 52.65.-y, 52.25.Dg, 52.25.-b, 52.65.Ff, 52.35.Tc
\end{description}
\end{abstract}
\maketitle

\paragraph{Introduction:}

\begin{figure*}
	\includegraphics[width=7in]{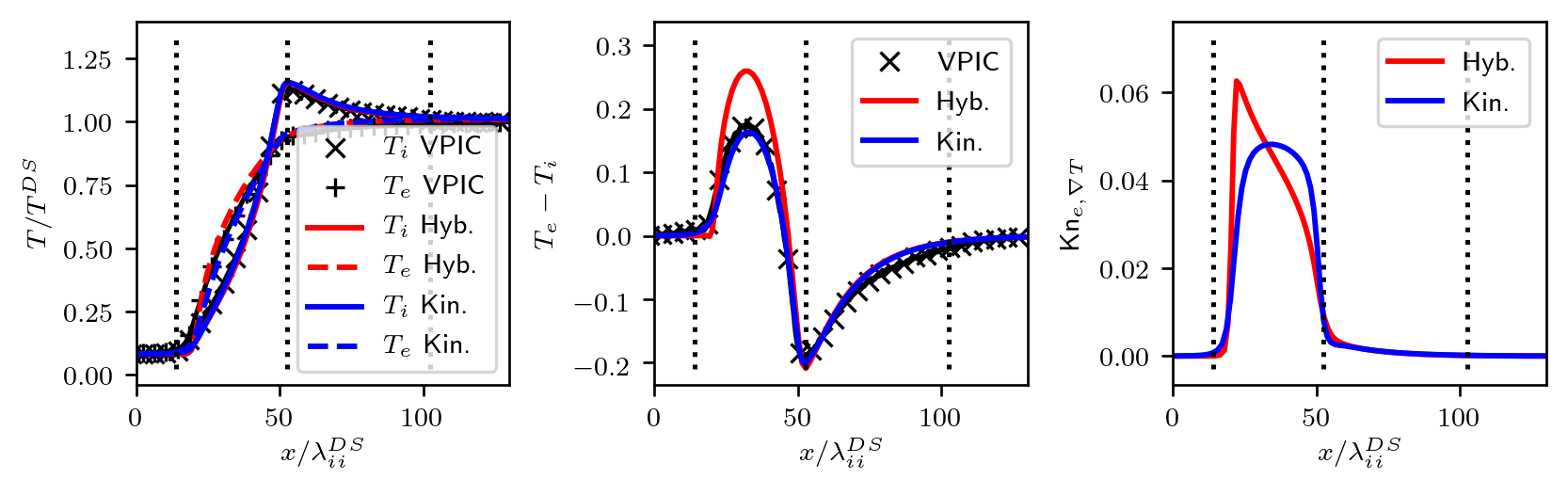}\caption{Left: Spatial profiles of $T_{i}$ (solid lines) and $T_{e}$ (dashed
		lines) normalized to the downstream limit for hybrid iFP (fluid electrons
		and kinetic ions, Hyb.), fully kinetic iFP (Kin.), and VPIC. Center:
		Spatial profiles of the temperature difference $T_{e}-T_{i}$ clearly
		showing a kinetic suppression of the temperature separation in the
		preheat layer. Right: Spatial profiles of $Kn_{e,\nabla T}$ for hybrid
		(Hyb.) and fully kinetic (Kin.) iFP simulations. \label{fig:temperature_profiles}}
\end{figure*}
Strong shocks are present in a variety of high-energy-density (HED)
environments, including inertial confinement fusion (ICF) capsule
implosions \citep{Rosenberg2015,Rosenberg2014a,Rinderknecht2014}.
Given the role of shock propagation in ICF compression and yield,
it is important to understand the impact that kinetic effects may
have in the shock structure, and its imprint on the imploding capsule
(e.g., Ref. \citep{Taitano2018b}). This is relevant, as the state-of-the-art
for simulating ICF capsule implosions is radiation-hydrodynamics (rad-hydro),
which is only strictly valid in systems with small Knudsen numbers
and therefore only accurate for weak shocks with $M\sim1$ \citep{Simakov2017a}.
Nevertheless, rad-hydro has been used extensively to investigate strong
plasma shocks \citep{Jukes1957b,Shafranov1957,Jaffrin1964,Grewal1973,Masser2011a,Zeldovich1967}.
The classical structure of a strong collisional plasma shock exhibits,
from upstream to downstream (see e.g., \citep{Zeldovich1967}): (1)
a prominent preheat layer wherein the electron temperature, $T_{e}$,
exceeds that of the ions, $T_{i}$, (2) an embedded ion compression
shock wherein $T_{i}$ increases rapidly and surpasses that of the
electrons, and (3) a region of ion-electron temperature equilibration.
The preheat and relaxation regions {[}(1) and (3){]} are approximately
of width $\approx Z_{i}^{2}\left(m_{i}/m_{e}\right)^{1/2}\lambda_{ii}^{DS}$
(where $m_{i}$, $m_{e}$, $Z_{i}$, and $\lambda_{ii}^{DS}$ are
the ion and electron masses, the ion charge, and the ion-ion downstream
mean-free-path, respectively), while the embedded compression shock
is of order several $\lambda_{ii}^{DS}$.

To date, kinetic studies of strong plasma shocks have focused on ions,
and have thus been performed almost exclusively with `hybrid' kinetic
codes employing a fluid-electron model coupled with a kinetic-ion
Vlasov-Fokker-Planck (VFP) description \citep{Casanova1991,Larroche1993,Vidal1993a,Keenan2017,Keenan2018}.
In these codes, the fluid electrons are quasi-neutral and ambipolar,
with $T_{e}$ determined from the electron energy equation and with
the electron heat flux modeled as the Braginskii/Spitzer-H\"{a}rm
expression (i.e., $\mathbf{\mathscr{Q}}_{Brag.,e}=-\kappa_{e}\nabla T_{e}$,
$\kappa_{e}\propto T_{e}^{5/2}$), usually with some variety of heat
flux limiter: $\left|\mathbf{\mathcal{\mathscr{Q}}}_{e}\right|=\min(f_{lim}\mathcal{\mathscr{Q}}_{FS,e},\left|\mathbf{\mathcal{\mathscr{Q}}}_{Brag.,e}\right|)$,
where $f_{lim}$ is effectively a `tuning knob' to prevent faster-than-streaming
diffusion, and $\mathcal{\mathscr{Q}}_{FS,e}=n_{e}T_{e}v_{th,e}$
is the free-streaming thermal flux. However, it is known that local
fluid models of the electron heat flux are only valid for sufficiently
small electron Knudsen numbers, (defined as the particle mean-free-path
over a characteristic gradient length scale, $Kn=\lambda/L$) $Kn_{e}\lesssim10^{-3}$.
For $Kn_{e}\gtrsim10^{-3}$, we expect to see deviations from the
fluid results. 

We note that sophisticated models for nonlocal electron heat flux
have been explored (such as the Luciani-Mora-Virmont (LMV) model \citep{Luciani1983,Bendib1988,Schurtz2000,Cao2015}).
Several of these models were investigated in Ref. \citep{Vidal1995}
with strong planar plasma shocks, using the kinetic ion code $\texttt{FPION}$
\citep{Casanova1991,Larroche1993,Vidal1993a}. However, while comparison
was made to a kinetic electron code \citep{Matte1982}, it was not
self-consistently coupled to the kinetic ions and utilized an expansion
in spherical harmonics, neglecting anisotropy in the collision operator.
Thus, to the authors' knowledge, a comparison of these nonlocal electron
heat flux models with fully kinetic self-consistent VFP simulations
is not yet available.

Recently, Zhang \emph{et al.} published results of fully kinetic particle-in-cell
simulations for an $M\approx3$ piston-driven planar plasma shock
\citep{Zhang2021}. Unfortunately, there are some significant limitations
in their work. Firstly, the shock has not detached from the piston;
secondly, the total simulation time is less than the ion-electron
thermal equilibration time \citep{Huba2013}; thirdly, the simulation
domain is roughly the shock width. Thus, their solution is transient
and has not yet relaxed to steady state. This is confirmed by the
fact that $T_{e}$ is non-monotonic and concave-upward in the preheat
layer {[}see their Fig. 4(b){]}, in significant contrast to what is
expected.

In this Letter, we present, for the first time, self-consistent simulations
of strong \emph{steady-state} collisional planar plasma shocks with
fully kinetic electrons and ions. We have performed these simulations
with two different fully kinetic codes (an Eulerian continuum and
a Lagrangian particle-in-cell), finding excellent agreement. The simulations
are performed with a realistic electron-ion mass ratio. We find that,
while kinetic electrons do not fundamentally change the structure
of the shock as compared with fluid electrons, they do modify appreciably
the structure of the preheat layer. In particular, the sharp electron-temperature
gradient at the front of the preheat layer disappears, and average
ion-electron temperature separation there decreases by about 20\%.
The rearrangement of the electron temperature profile in the preheat
layer can be traced to nonlocal electron heat transport effects due
to the kinetic nature of electrons. Moreover, we find that the electron
heat-flux profile displays \textquotedblleft stiffness\textquotedblright{}
in that it qualitatively agrees remarkably well between fluid and
kinetic electron models, demanding adjustments in the corresponding
electron temperature profile. We also verify the LMV nonlocal electron
heat-flux closures proposed in the literature against self-consistent
full VFP simulations (to the authors' knowledge, the first time this
has been done), and find them to provide a reasonable agreement with
the high-fidelity simulations.

\paragraph{Problem setup:}

To study this problem, we employ the Eulerian VFP code iFP \citep{Anderson2020,Taitano2018,Taitano2021,Taitano2021b},
and the Lagrangian code VPIC \citep{Bowers2008b,Bowers2008c,Bowers2009}.
The iFP code solves the coupled VFP equations for each of the plasma
species in a 1D planar electrostatic approximation and the electric
field is given by the 1D Amp\`{e}re equation:
\begin{align}
\partial_{t}\left(f_{\alpha}\right)+\partial_{x}\left(v_{\parallel}f_{\alpha}\right)+\frac{Z_{\alpha}e}{m_{\alpha}}E_{\parallel}\partial_{v_{\parallel}}\left(f_{\alpha}\right) & =\sum_{\beta=1}^{N_{s}}C_{\alpha\beta},\label{eq:VFP}\\
\epsilon_{0}\partial_{t}\left(E_{\parallel}\right)+\sum_{\alpha=1}^{N_{s}}Z_{\alpha}enu_{\parallel,\alpha} & =0,\label{eq:Ampere}
\end{align}
where $\alpha,\beta$ denote the species index, $f_{\alpha}$ is the
particle distribution function for each species; the symbols $Z_{\alpha}$
and $m_{\alpha}$ indicate each species' particle charge and mass,
and $e$ is the proton charge. The electric field and $\alpha$-species
bulk velocity in the shock propagation direction are denoted by $E_{\parallel}$
and $u_{\parallel,\alpha}$, respectively. We denote vector components
along the spatial direction of the shock propagation ($x$) with the
subscript $_{\parallel}$, while perpendicular components are denoted
with the subscript $_{\bot}$. The VPIC code is a 3D electromagnetic
particle-in-cell code, though for the simulations presented here,
it has been limited to 1D in configuration-space.

We consider planar hydrogen plasma shocks with protons \emph{p}\textsuperscript{\emph{+}}
and electrons \emph{e}\textsuperscript{\emph{-}}, with the realistic
mass ratio $m_{p}/m_{e}=1836$. The computations are performed in
the frame of the shock. We initialize the computations with the Rankine-Hugoniot
relations as upstream and downstream boundary conditions, with a hyperbolic
tangent transition in between. In order to allow room for the downstream
equilibration zone, the compression-shock position is off-centered,
at $x/L=0.4$, where $L$ is the total domain length. Unless otherwise
specified, the total simulation domain is 250 downstream ion-ion mean-free-paths,
i.e. $L=250\lambda_{ii}^{DS}$. In the regime considered here, the
Debye length $\lambda_{D}$ is much smaller than collisional mean-free-paths,
$\lambda_{D}\ll\lambda_{ii}$, everywhere.

The VPIC simulations were initialized by sampling from the steady-state
iFP ion and electron distribution functions over a truncated spatial
domain ($L=130\lambda_{ii}^{DS})$ and run to steady state. To mitigate
the noise inherent in the particle-in-cell simulations, we have applied
a Savitzky-Golay filter \citep{SavitzkyA.;Golay1964} to smooth the
resulting VPIC moment profiles. In all the figures, the vertical dotted
lines indicate (from left to right) the approximate locations of the
upstream edge of the preheat layer, the compression shock, and the
downstream edge of the relaxation layer. The spatial profiles presented
here are truncated to a span of $130\lambda_{ii}^{DS}$ to match the
VPIC simulation domain.

\paragraph*{Results:}

We consider a strong, $M=6$ shock. In Fig. \ref{fig:temperature_profiles}-left
and -center, we see the spatial profiles of temperature and temperature
difference ($T_{e}-T_{i}$) for ions and electrons comparing hybrid
(kinetic ions and fluid electrons) to fully kinetic simulations. The
profiles of the bulk velocity and density moments do not change appreciably
between hybrid and fully kinetic simulations.  In these moment profiles,
the main difference between hybrid and fully kinetic results is the
suppression of the $\delta T=T_{e}-T_{i}$ in the preheat layer ($x\sim15$
\textendash{} $55$) by approximately $\sim20\%$, with $T_{e}$ relaxing
towards $T_{i}$ (see. Fig. \ref{fig:temperature_profiles}-center).
The temperature profiles through the compression shock and into the
downstream equilibration layer are essentially unchanged between the
fluid and kinetic electron treatments. 

In Fig. \ref{fig:temperature_profiles}-right we compare the electron
Knudsen number based upon the temperature-gradient length scale {[}$Kn_{e,\nabla T}\equiv\lambda_{e}\left|\nabla\left(\ln T_{e}\right)\right|${]}
for fully kinetic and hybrid iFP simulations.  We see that $Kn_{e}$
is approximately $5\times10^{-2}$ throughout the preheat layer (well
above the $Kn_{e}\sim10^{-3}$ threshold), which corresponds exactly
to where the electron-temperature profile has been adjusted in Fig.
\ref{fig:temperature_profiles}, and is much smaller elsewhere. Notably
the sharp peak in fluid-electron $Kn_{e}$ profile at the preheat
layer upstream edge (where $\partial_{x}T_{e}$ is largest) is smoothed
significantly in the kinetic-electron case. 

\begin{figure}
	\includegraphics[width=3.375in]{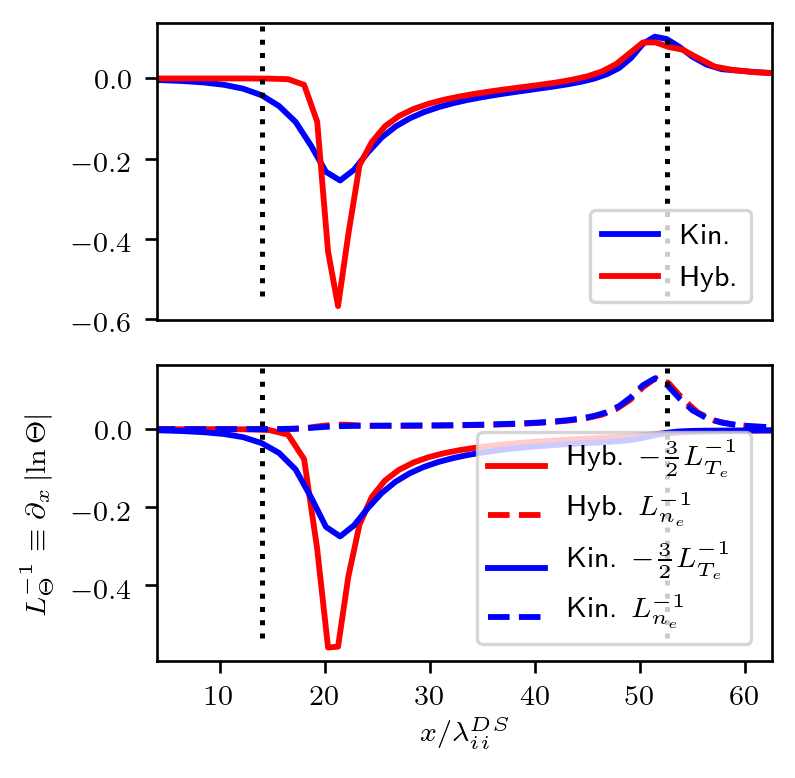}\caption{Top: Spatial profiles within the preheat region for the $\partial_{x}\mathscr{Q}_{\parallel,e}/T_{e}$
		term in Eq. \eqref{eq:temperature-alt} for hybrid (Hyb.) and fully-kinetic
		(Kin.) simulations. Bottom: Spatial profiles of the inverse gradient
		length scales based on $T_{e}$ (solid) and $n_{e}$ (dashed) for
		hybrid (Hyb.) and fully-kinetic (Kin.) simulations. \label{fig:energy_balance}}
\end{figure}
\begin{figure*}
	\includegraphics[width=7in]{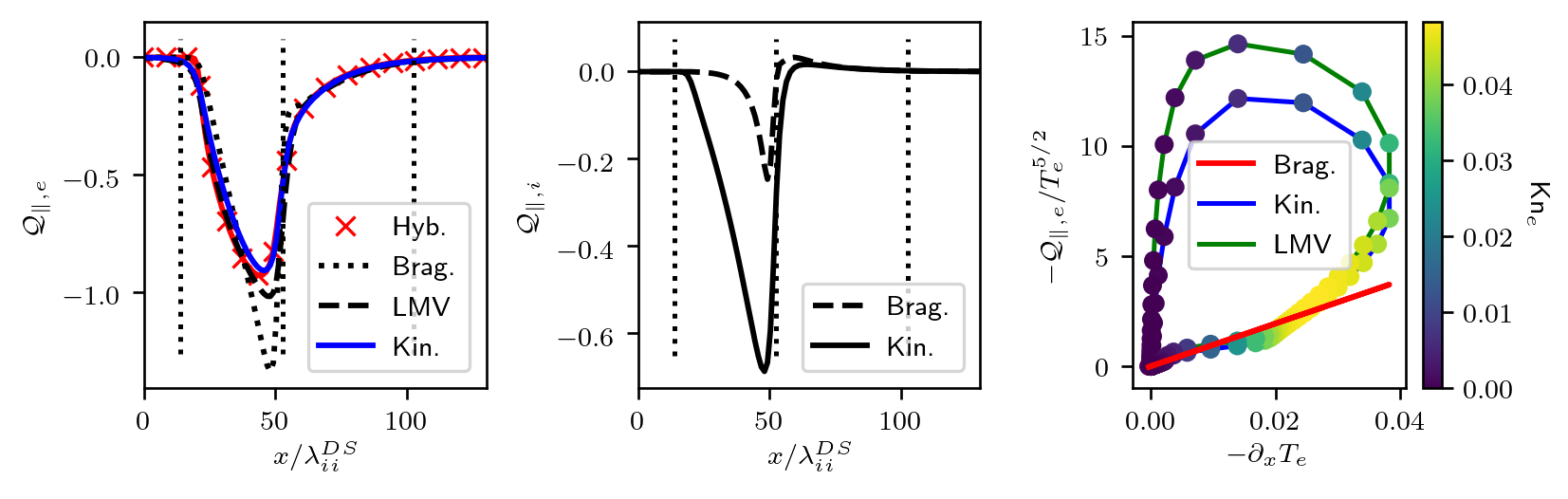}
	
	\caption{Left: Spatial profiles of the electron heat flux comparing $\mathscr{Q}_{\parallel,e}^{hyb.}$
		(Hyb.), $\mathscr{Q}_{\parallel,e}^{kin.}$ (Kin.), and $\mathscr{Q}_{\parallel,e,Brag.}^{kin.}$
		(Brag.) and $\mathscr{Q}_{\parallel,e,LMV}^{kin.}$ (LMV) heat fluxes
		from the fully kinetic electron moments. Note the similarity of $\mathscr{Q}_{\parallel,e}^{hyb.}$
		and $\mathscr{Q}_{\parallel,e}^{kin.}$. Center: Spatial profiles
		of $\mathscr{Q}_{\parallel,i}^{kin.}$ (Kin.) and $\mathscr{Q}_{\parallel,e,Brag.}^{kin.}$
		(Brag.). Note $\mathscr{Q}_{\parallel,i}^{kin.}$ is comparable in
		magnitude to $\mathscr{Q}_{\parallel,e}^{kin.}$. Right: Phase-space
		plot of the electron heat flux comparing $\mathscr{Q}_{\parallel,e}^{kin.}$
		(Kin.) to the $\mathscr{Q}_{\parallel,e}^{hyb.}$ (Hyb.) and the $\mathscr{Q}_{\parallel,e,LMV}^{kin.}$
		(LMV). The heat flux curves for $\mathscr{Q}_{\parallel,e}^{kin.}$
		and $\mathscr{Q}_{\parallel,e,LMV}^{kin.}$ have symbols colored by
		local Knudsen numbers. \label{fig:heatflux_space}}
\end{figure*}
To look deeper into differences between hybrid and fully kinetic simulations,
we examine the balance of terms in the electron energy equation (see
Eq. (A3) of Ref. \citep{Simakov2016}). Our simulations are performed
in the frame of the shock and allowed to reach a steady state, so
the temporal term is neglected. Further, our analysis showed the contribution
of the electron viscosity, friction, and thermal relaxation terms
to be negligible. Upon rearranging, we find the energy equation may
be expressed as 
\begin{equation}
\frac{\partial_{x}\mathscr{Q}_{\parallel,e}}{T_{e}}\approx n_{e}u_{\parallel,e}\left[\partial_{x}\left(\ln n_{e}\right)-\frac{3}{2}\partial_{x}\left(\ln T_{e}\right)\right].\label{eq:temperature-alt}
\end{equation}
As the particle flux density $n_{e}u_{\parallel,e}$ is constant through
the shock, Eq. \eqref{eq:temperature-alt} clearly expresses that
the ratio of the divergence of the electron heat flux to the electron
temperature is dependent only on the two inverse length scales, $L_{T_{e}}^{-1}\equiv\left|\partial_{x}\left(\ln T_{e}\right)\right|$
and $L_{n_{e}}^{-1}\equiv\left|\partial_{x}\left(\ln n_{e}\right)\right|$.
This is demonstrated in Fig. \ref{fig:energy_balance}, which shows
the left-hand side of Eq. \eqref{eq:temperature-alt} (Top) and the
inverse gradient length scales for the electron-temperature and electron-density
profiles (Bottom). From this, we can clearly conclude that the primary
effect of the kinetic electrons is to smooth the steep electron-temperature
gradient at the preheat layer edge where the Knudsen number (based
on $T_{e}$) is the largest.

It follows from Fig. \ref{fig:temperature_profiles}-right and Fig.
\ref{fig:energy_balance} that, in the preheat layer, $T_{e}^{Kin.}\sim T_{e}^{Hyb.}$
and $L_{T_{e}}^{Kin.}\sim2L_{T_{e}}^{Hyb.}$. Then, Eq. \eqref{eq:temperature-alt}
predicts in this region $\mathscr{Q}_{\parallel,e}^{Kin.}\approx\mathscr{Q}_{\parallel,e}^{Hyb.}$.
Figure \ref{fig:heatflux_space}-left demonstrates this quite clearly,
where we see spatial profiles of the kinetic electron parallel heat
flux, $\mathscr{Q}_{\parallel,e}^{kin.}$, compared to the Braginskii
heat flux in the hybrid simulation, $\mathscr{Q}_{\parallel,e}^{hyb.}$.
Here, we compute the kinetic heat flux for a species $\alpha$ from
its distribution function $f_{\alpha}$ using 
\begin{equation}
\mathscr{Q}_{\parallel,\alpha}^{kin.}=\int_{\bm{v}}\frac{1}{2}m_{\alpha}\left(\bm{v}-\bm{u}_{\alpha}\right)^{2}\left(v_{\parallel}-u_{\parallel,\alpha}\right)f_{\alpha}d\bm{v},\label{eq:heat_flux}
\end{equation}
where $\bm{u}_{\alpha}$ is the $\alpha$-species bulk velocity. For
comparison, we also compute from the kinetic-electron moments a Braginskii
heat flux, $\mathscr{Q}_{\parallel,e,Brag.}^{kin.}$, and a heat flux
based from the kinetic closure LMV model (with electric-field correction
\citep{Bendib1988,Schurtz2000}), $\mathscr{Q}_{\parallel,e,LMV}^{kin.}$.
The LMV model is a nonlocal closure of the form
\begin{equation}
\mathscr{Q}_{\parallel,LMV}=\int_{-\infty}^{+\infty}W\left(x,x'\right)\mathscr{Q}_{\parallel,Brag.}\left(x'\right)\frac{dx'}{a\lambda_{e}\left(x'\right)},\label{eq:LMV}
\end{equation}
where $W\left(x,x'\right)$ is a phenomenologically-chosen delocalization
kernel for the Braginskii heat flux, $\lambda_{e}\left(x'\right)$
is an appropriate delocalization length scale (related to the mean-free-path
and potentially containing an electric field correction), and $a$
is an adjustable parameter (Ref. \citep{Schurtz2000} recommends $a=32$).
We note that, as we expected from the analysis of Eq. \eqref{eq:temperature-alt}
and Fig. \ref{fig:energy_balance}, the kinetic and hybrid heat fluxes
$\mathscr{Q}_{\parallel,e}^{kin.}$ and $\mathscr{Q}_{\parallel,e}^{hyb.}$
are very similar in overall magnitude and shape. The difference in
the profile is essentially in a `smoothing' of the gradients at the
upstream edge of the preheat layer (essentially the same modification
as seen in the electron-temperature profile). For the post-processed
heat-flux models based on the kinetic electron-moment profiles, we
see that the LMV heat flux, $\mathscr{Q}_{\parallel,e,LMV}^{kin.}$,
does very well, while $\mathscr{Q}_{\parallel,e,Brag.}^{kin.}$ is
quite different. For reference, the kinetic ion heat flux, $\mathscr{Q}_{\parallel,i}^{kin.}$,
and the corresponding Braginskii heat flux, $\mathscr{Q}_{\parallel,i,Brag.}^{kin.}$
are also included in Fig. \ref{fig:heatflux_space}-center. There,
we observe that $\mathscr{Q}_{\parallel,i}^{kin.}$ is of the same
order of magnitude as $\mathscr{Q}_{\parallel,e}^{kin.}$ and is substantially
larger than $\mathscr{Q}_{\parallel,i,Brag.}^{kin.}$. 

Despite the similarity between hybrid and kinetic-electron heat fluxes,
a detailed analysis reveals deeper physical differences. In particular,
we find clear evidence of nonlocal transport effects in the kinetic-electron
simulation, as expected from weakly collisional conditions. This is
evidenced in Fig. \ref{fig:heatflux_space}-right, where we show a
phase-space plot of the heat flux vs. $\partial_{x}T_{e}$, with symbols
colored according to the local Knudsen number. We see that the kinetic
heat flux demonstrates a significant departure from the Braginskii
model, with the nonlocal kinetic heat flux possessing a distinct multi-valued
dependence on the temperature gradient. Notably, the LMV heat-flux
model also demonstrates a\emph{ }very similar nonlocal behavior, though
it appears to over-predict slightly the nonlocality relative to the
true kinetic heat flux. We note the LMV without electric-field correction
(i.e., the original model of Luciani \emph{et al.} \citep{Luciani1983},
not shown here) over-predicts the heat flux by as much as a factor
of 2 (the peak in Fig. \ref{fig:heatflux_space}-right would be nearer
to $-\mathscr{Q}_{\parallel,e}/T_{e}^{5/2}=30$), though with a similar
qualitative shape.
\begin{figure}
	\includegraphics[width=3.375in]{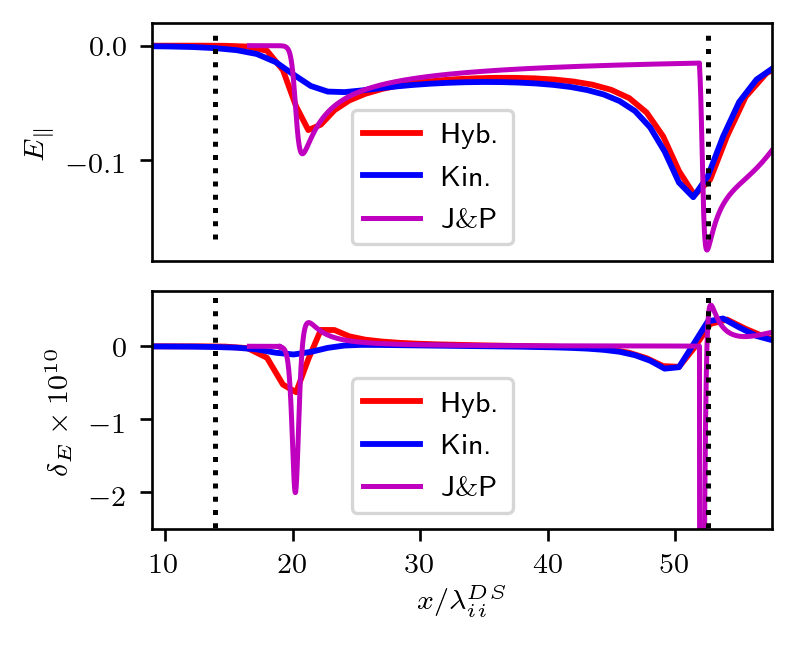}\caption{Top: Spatial profiles of $E_{\parallel}$ from hybrid (Hyb.) and fully
		kinetic iFP (Kin.) simulations. Also included is the estimated electric
		field from the semi-analytic solution of Ref. \citep{Jaffrin1964}
		(J\&P). Bottom: Spatial profiles of the charge separation quantity
		$\delta_{E}$, again comparing hybrid and fully kinetic iFP and the
		semi-analytic estimate of Ref. \citep{Jaffrin1964} \label{fig:E_r_plot}}
\end{figure}

The electric field, $E_{\parallel}$, shows further differences between
hybrid and fully kinetic simulations. In the fully kinetic simulation,
$E_{\parallel}$ is solved consistently from Eq. \ref{eq:Ampere},
while in the hybrid simulation it is obtained from the electron momentum
equation by neglecting electron inertia, viscosity, and the electron-ion
friction term, 
\begin{equation}
E_{\parallel}=\frac{1}{q_{e}n_{e}}\partial_{x}\left[n_{e}T_{e}\right].\label{eq:momentum}
\end{equation}
In Fig. \ref{fig:E_r_plot}-top, we see the spatial profiles of
$E_{\parallel}$ for hybrid and fully kinetic iFP simulations. As
expected, the most noticeable difference is in the preheat layer,
where the electron-temperature gradient has been significantly reduced,
suppressing the $E_{\parallel}$ `spike' there. This preheat layer
spike corresponds to the 'precursor' electric shock layer described
by \citeauthor{Jaffrin1964} \citep{Jaffrin1964}. Included in Fig.
\ref{fig:E_r_plot}-top is a profile based on their semi-analytic
shock solution. We note that the derivation of Ref. \citep{Jaffrin1964}
is based on a Navier-Stokes model. To make a more fair quantitative
comparison, we have altered the solution by making the transformations
$\epsilon\rightarrow\epsilon/2$ in their Eq. (3.4a) and $\epsilon\rightarrow2\epsilon$
in their Eq. (3.5a) (to bring the ratio of electron and ion transport
coefficients more into line with the estimates of \citeauthor{Braginskii1965}
\citep{Braginskii1965}). The result shows remarkably good agreement
with the hybrid iFP solution in the precursor shock layer at the upstream
preheat layer edge, but agrees only qualitatively elsewhere. We include
an analysis of the charge separation, $\delta_E =  \epsilon_0 \partial_x E_\parallel $, in Fig. \ref{fig:E_r_plot}-bottom.
Given that $\lambda_{D}/\lambda_{ii}\ll1$, we see that, unsurprisingly,
the charge separation for this problem is quite small. We also again
see the precursor shock collapse in the fully kinetic case. However,
the charge separation predicted by the semi-analytic solution in the
preheat layer is within an order of magnitude of that shown by the
hybrid simulation. Reference \citep{Jaffrin1964} estimates the precursor
shock thickness to be $l\sim M\lambda_{ii}^{US}/\sqrt{m_{e}/m_{i}}$
(here $l\sim6$), which is about a factor of two smaller than the
width estimated from the hybrid and fully kinetic simulations.

\paragraph*{Conclusions:}

We have performed the first converged fully kinetic simulations of
strong steady-state planar plasma shocks. We have found that the differences
between hybrid and fully kinetic simulations are limited to the electron
preheat region of the shock, and focused at the upstream edge, where
the temperature gradient length scale is smallest and the electron
temperature Knudsen number is largest. Due to its strong connection
to the electron-temperature profile {[}through Eq. \eqref{eq:temperature-alt}{]},
we find the electron heat flux exhibits stiffness across various models,
with only small adjustments to the heat flux slope near the upstream
edge of the preheat layer. However, while its spatial profile is qualitatively
unchanged, the heat flux exhibits significant nonlocality in the fully
kinetic case. We find that a heat flux computed using the LMV model
\citep{Luciani1983,Bendib1988,Schurtz2000} using kinetic-electron
moments recaptures the kinetic-electron heat flux reasonably well
including the nonlocal phase-space behavior, while the corresponding
Braginskii heat flux shows significant differences. The electric field,
$E_{\parallel}$, at the preheat layer leading edge is found to be
very similar to that predicted by a semi-analytic fluid solution \citep{Jaffrin1964},
but with minimal charge separation due to the gradient smoothing effect
of the kinetic electrons. 

\begin{acknowledgments}
	This research was performed under the auspices of the National Nuclear
	Security Administration of the U.S. Department of Energy at Los Alamos
	National Laboratory (LANL) under Contract No. 89233218CNA000001. And
	was supported by funding from the Advanced Simulation and Computing
	Program (ASC) Thermonuclear Burn Initiative (TBI) and Laboratory Directed
	Research and Development (LDRD), with the use of LANL Institutional
	Computing (IC) resources. The authors thank Dr. Bill Daughton for
	assistance with setting up the VPIC particle-in-cell simulations.
\end{acknowledgments}

\bibliographystyle{aapmrev4-2}
%\bibliography{library}
%

\end{document}